\documentclass{article}
\usepackage{kr}
\pdfoutput=1

\usepackage{listings}
\usepackage{amsthm}

\newcommand {\ri} {\rightarrow}
\newcommand {\Ri} {\Rightarrow}

\newcommand {\Unt} {{\cal U}}

\newcommand {\ent} {\mathrel{{\scriptstyle\mid\!\sim}}}

\newcommand {\la} {\langle}
\newcommand {\ra} {\rangle}
\newcommand {\sx} {\langle}
\newcommand {\dx} {\rangle}
\newcommand {\emme} {\mathcal{M}}
\newcommand {\cali} {\mathcal{I}}

\newcommand {\WW} {\mathcal{W}}

\newcommand{\tip}{{\bf T}}

\newcommand{\alc}{\mathcal{ALC}}

\newcommand{\be}{\begin{enumerate}}
\newcommand{\ee}{\end{enumerate}}

\newcommand{\hide}[1]{}

\usepackage{times}
\usepackage{soul}
\usepackage{url}
 \usepackage[hidelinks]{hyperref}
 \usepackage[utf8]{inputenc}
 \usepackage[small]{caption}
 \usepackage{graphicx}
 \usepackage{amsmath}
\usepackage{amsthm}
 \usepackage{booktabs}
 \usepackage{algorithm}
 \usepackage{algorithmic}
\usepackage{times}
\usepackage{graphicx}
\usepackage{latexsym}

\usepackage{multicol,lipsum}
\usepackage{microtype}
\usepackage{amssymb}
\usepackage{color}
\usepackage{ifsym} 

\newtheorem{proposition}{Proposition}
\newtheorem{definition}{Definition}

\title{Temporal  Many-valued Conditional Logics: \\
a Preliminary Report
}

\author{Mario Alviano$^1$\and
Laura Giordano$^2$\and
Daniele Theseider Dupr{\'{e}}$^2$\\
\affiliations
$^1$Universit\`a della Calabria,  Italy\\
$^2$Universit\`a del Piemonte Orientale, Italy \\
\emails
alviano@mat.unical.it,
\{laura.giordano,dtd\}@uniupo.it
}

\begin{document}

\maketitle

\begin{abstract}
In this paper we propose a many-valued temporal conditional logic. We start from a many-valued logic with typicality, and extend it with the temporal operators of the Linear Time Temporal Logic (LTL), thus providing a formalism which is able to capture the dynamics of a system, trough strict and  defeasible temporal properties. 
We also consider an instantiation of the formalism for gradual argumentation.
\end{abstract}

\section{Introduction}

Preferential approaches to  commonsense reasoning 
\cite{Delgrande:87,Makinson88,Pearl:88,KrausLehmannMagidor:90,Pearl90,whatdoes,BenferhatIJCAI93,BoothParis98,Kern-Isberner01} 
have their roots in conditional logics \cite{Lewis:73,Nute80},
and  have been used to provide axiomatic foundations of non-monotonic or defeasible reasoning.

In recent work  \cite{NMR2023}, we have proposed a many-valued multi-preferential conditional logic with typicality to define a preferential interpretation of an argumentation graph in gradual argumentation semantics 
\cite{CayrolJAIR2005,Dunne2011,Amgoud2017,BaroniRagoToni2018,BaroniRT19,Amgoud2019}.

This paper aims at  defining a {\em propositional many-valued temporal logic with typicality}, by extending the many-valued conditional logic with typicality developed in \cite{NMR2023} with temporal operators from the Linear Time Temporal Logic (LTL). This allows considering the temporal dimension, when reasoning about the  defeasible typicality properties of a system, for explanation, such as by capturing the dynamics of a weighted Knowledge Base (KB) \cite{IJAR23}.

Preferential extensions of LTL with defeasible temporal operators have been recently studied  \cite{ChafikACV21,ChafikTIME2020,ChafikThesis} to enrich temporal formalisms with non-monotonic reasoning features, by considering defeasible versions of the LTL operators.
Our approach, instead, will consist in adding the  standard LTL operators to a (many-valued) conditional logic with typicality,
an approach similar to the preferential extension considered for Description Logics (DLs)  in \cite{CILC2023}, where the logic $\mathit{LTL}_{\alc}$ \cite{Lutz08}, extending $\alc$ with LTL operators, has been further extended with a {\em typicality operator}, to develop a (two-valued) temporal $\alc$ with typicality, $\mathit{LTL}_{\alc}^\tip$. 

As in the Propositional Typicality Logic by Booth et al.\  \cite{BoothCasiniAIJ19} (and in the DLs with typicality  \cite{FI09}) the conditionals are formalized based on material implication (resp., concept inclusion in DLs) plus the {\em typicality operator} $\tip$.
The typicality operator allows for the definition of {\em conditional implications} 
$\tip(\alpha) \rightarrow  \beta$, meaning that ``normally if $\alpha$ holds,  $\beta$ holds".
They correspond to conditional implications $\alpha \ent \beta$ in  KLM logics \cite{KrausLehmannMagidor:90,whatdoes}. More precisely in this paper, as in \cite{NMR2023}, we consider a many-valued semantics, so that a formula is given a value in a {\em truth degree set} ${\cal D}$, and the two-valued case can be regarded as a specific case, obtained for ${\cal D}=\{0, 1\}$.
As the logic is many valued, we consider  {\em graded conditionals} 
of the form \linebreak $\tip(\alpha) \rightarrow  \beta \geq l$ (resp., $\tip(\alpha) \rightarrow  \beta \leq l$), meaning that  ``normally if $\alpha$ holds then  $\beta$ holds with degree at least (resp., at most) $l$".
For instance, the formalism allows for representing graded implications such as:
$$ \mathit{living}\_\mathit{in}\_\mathit{Town} \wedge \mathit{Young} \ri \tip(\Diamond \mathit{Granted\_Loan}) \geq l,$$
meaning that living in town and being young, implies that normally the loan is eventually granted with degree at least $l$, where the interpretation of some concepts  (e.g., $\mathit{Young}$) may be non-crisp.

The preferential semantics of the logic exploits {\em multiple preference relations} $<_\alpha$ with respect to different formulas $\alpha$, following the approach developed for ranked and weighted KBs in description logics, based on a {\em multi-preferential semantics} \cite{TPLP2020,JELIA2023} 
and for conditionals in the propositional calculus in \cite{AIJ21}, where preference are allowed with respect to different aspects.

The schedule of the paper is the following. Section  \ref{sec:many_valued_LTL} develops a many-valued preferential logic with typicality, inspired to 
 \cite{NMR2023} (but not specifically intended for argumentation).  Section \ref{sec:Conditional_reasoning} extends such logic with LTL modalities to develop a temporal many-valued conditional logic, and {\em temporal graded formulas}.
In Section \ref{sec:weighted_KBs}, we introduce weighted temporal knowledge bases and their semantics. In Section \ref{sec:gradual_argumentation}, we consider an instantiation of the logic for gradual argumentation, in the direction of providing a temporal conditional semantics for reasoning about the dynamics of gradual argumentation graphs. Section \ref{sec:conclusions} concludes the paper.


\section{A Many-valued Preferential  Logics with Typicality} \label{sec:many_valued_LTL}

In this section we define a many-valued propositional logic with typicality.

Let ${\cal L}$ be a propositional many-valued logic, whose formulas are built from a set $Prop$ of propositional variables using the logical connectives $\wedge$, $\vee$, $\neg$ and $\rightarrow$, as usual. 
We assume that $\bot$ (representing falsity) and $\top$ (representing truth) are formulas of ${\cal L}$.
We consider a many-valued semantics for formulas, over a {\em truth degree set} ${\cal D}$,
equipped with a preorder relation $\leq^{\cal D}$,
a bottom element $0^{\cal D}$, and a top element $1^{\cal D}$. 
 We denote by $<^{\cal D}$ and $\sim^{\cal D}$ the related strict preference relation and equivalence relation (often, we will omit explicitly referring to ${\cal D}$, and simply write $\leq$ $<$, $\sim$,  $0$ and $1$).

Let $\otimes$, $\oplus$, $\ominus$ and $\rhd$ be 
the {\em truth degree functions} in ${\cal D}$ for the connectives $\wedge$, $\vee$, $\neg$ and $\rightarrow$ (respectively).
When  ${\cal D}$ is $[0,1]$ or the finite truth space
${\cal C}_n= \{0, \frac{1}{n},\ldots,$ $ \frac{n-1}{n}, \frac{n}{n}\}$, for an integer $n \geq 1$, as in our case of study \cite{ASPOCP23}, $\otimes$, $\oplus$, $\rhd$ and $\ominus$ can be chosen as a t-norm, an s-norm, an implication function, and a negation function
in some system of many-valued logic \cite{Gottwald2001};
for instance, in G\"odel logic (that we will consider later): $a \otimes b= min\{a,b\}$,  $a \oplus b= max\{a,b\}$,  $a \rhd b= 1$ {\em if} $a \leq b$ {\em and} $b$ {\em otherwise}; and  $\ominus a  = 1$ {\em if} $a=0$ {\em and} $0$ {\em otherwise}.

We further extend the language of ${\cal L}$ by adding a typicality operator as introduced by Booth et al. \cite{BoothCasiniAIJ19} for propositional calculus, 
and by Giordano et al.\ for preferential description logics \cite{lpar2007}. 
Intuitively, ``a sentence of the form $\tip(\alpha)$ is understood to refer to the {\em typical situations in which $\alpha$ holds}" \cite{BoothCasiniAIJ19}.
The typicality operator allows the formulation of  {\em  conditional implications} (or {\em defeasible implications}) of the form 
$\tip(\alpha) \rightarrow \beta$
whose meaning is that ``normally, if $\alpha$ then $\beta$'', 
or ``in the typical situations when $\alpha$ holds, $\beta$ also holds''.
They correspond to conditional implications $\alpha \ent \beta$ of KLM preferential logics \cite{whatdoes}.
As in PTL \cite{BoothCasiniAIJ19}, the typicality operator cannot be nested.
When $\alpha$ and $\beta$ do not contain occurrences of the typicality operator, an implication $\alpha \rightarrow \beta$ is called {\em strict}. 
We call ${\cal L}^\tip$ the language obtained by extending  ${\cal L}$ with a unary typicality operator $\tip$. 
In the logic ${\cal L}^\tip$, we allow  {\em general implications}  $\alpha \rightarrow \beta$, 
where $\alpha$ and $\beta$ may contain occurrences of the typicality operator.

The interpretation of a typicality formula  $\tip(\alpha)$ is defined with respect to a preferential interpretation.
The KLM preferential semantics \cite{KrausLehmannMagidor:90,whatdoes,Pearl:88} exploits a set of worlds $\WW$, with their valuation and a preference relation $<$ among worlds, to provide an interpretation of conditional formulas. A conditional $A\ent B$ is  satisfied
in a preferential interpretation, if $B$ holds in all the most normal worlds satisfying $A$, i.e., in all $<$-minimal worlds satisfying $A$.

Here we consider a many-valued multi-preferential semantics. The propositions at each world $w \in \WW$ have a value in ${\cal D}$ and multiple preference relations $<_{A} \subseteq W\times W$ are associated to formulas $A$ of ${\cal L}$.

Multi-preferential semantics have been previously considered for defining refinements of the rational closure construction \cite{AIJ21,GliozziAIIA2016}, as well as for defeasible DLs, both in the two-valued case (e.g., for ranked defeasible KBs \cite{iclp2020}), and in the many-valued case (e.g., for weighted conditional KBs \cite{JELIA2021,IJAR23}).
The semantics below exploits a set of preference relations $<_{A_i}$ associated to the formulas $A_i$ of ${\cal L}$.

\begin{definition}\label{MPinterpretations}
A {\em (multi-)preferential  interpretation}  is a triple $\emme= \sx \WW, \{<_{A_i}\}, v \dx$ where:
\begin{itemize} 
\item $\WW$ is a non-empty set of worlds;
\item  each $<_{A_i} \subseteq \WW \times \WW$ is an irreflexive and transitive relation on $\WW$;
\item  $v: \WW \times \mathit{Prop} \longrightarrow {\cal D}$ is a valuation function, assigning a truth value in ${\cal D}$ to any propositional variable in each world $w \in \WW$. 
\end{itemize}
\end{definition} 
\noindent
The valuation $v$ is inductively extended to  all formulas in ${\cal L}^\tip$: \begin{quote}
$v(w,\bot)=0_{\cal D}$ \ \ \ \ \ \ \ \ \  $v(w,\top)=1_{\cal D}$

  $v(w, A \wedge B)= v(w,A) \otimes v(w,B)$
  
  $v(w,A \vee B)= v(w,A) \oplus v(w,B)$ 
  
  $v(w,A \rightarrow B)= v(w,A) \rhd v(w,B)$
     
  $v(w,\neg A)= \ominus v(w,A)$
\end{quote} 
 \noindent
and the interpretation of a typicality formula  $\tip(A)$  in $\emme$, at a world $w$, is defined as:  
\begin{align*}\label{eq:interpr_typicality}
	v(w, \tip(A))  = \left\{\begin{array}{ll}
						 v(w, A) & \mbox{if } \mathit{\nexists w' \in \WW} \ \mbox{s.t. } \mathit{ w' <_A w} \\
						0_{\cal D} &  \mbox{otherwise } 
					\end{array}\right.
\end{align*}  
When $v(w, \tip(A))\neq 0_{\cal D}$, $w$ is a typical/normal $A$-world in $\emme$. Note that we do not assume well-foundedness of $<_A$.

A $\mathit{ranked}$ interpretation is a (multi-)preferential interpretation  $\emme= \sx \WW,\{<_{A_i}\}, v \dx$ for which the preference relations $<_A$ are modular, that is: for all $x, y, z$, if
$x < _Ay$ then $x <_A z$ or $z <_A y$.

We can now define the satisfiability in $\emme$ of a {\em graded implication}, with form $A  \rightarrow B \geq l$ or $A  \rightarrow B \leq u$, where $l$ and $u$ are constants corresponding to truth values in ${\cal D}$ and $A$ and $B$ are formulas of ${\cal L}^\tip$.

Given a preferential interpretation $\emme= \sx \WW, \{<_{A_i}\}, v \dx$,  
we can define the truth degree of an implication $A \rightarrow B$ in  $\emme$ as follows:
\begin{definition}
Given a preferential interpretation  $\emme= \sx \WW, \{<_{A_i}\}, v \dx$  
the {\em truth degree of an implication $A \rightarrow B$ wrt. $\emme$} is defined as: \\ 
$\mbox{\ \ \ }$ \ \ \ $(A \rightarrow B)^\emme= inf_{w \in \WW} (v(w,A) \rhd v(w, B) )$.
\end{definition}

In general, some conditions may be needed to enforce an {\em agreement} between the truth values of a formula $A$ at the different worlds in $\emme$ and preference relations $<_A$ among them. The preferences $<_A$ might have been determined by some {\em closure construction}, such as those exploiting the ranks or weights of conditionals, as in \cite{iclp2020,JELIA2021}.
Similar conditions, called coherence, faithfulness and  $\varphi$-coherence conditions, have for instance been introduced in the multi-preferential semantics for DLs with typicality in \cite{JELIA2021,IJAR23}.  
Below we introduce a  {\em coherence}  and a {\em faithfulness} condition.

We call a (multi-)preferential interpretation $\emme= \sx \WW, \{<_{A_i}\}, v \dx$  {\em coherent} if, for all $w,w' \in \WW$, and preference relation $<_{A_i}$, 
$$ v(w,A_i) > v(w',A_i)  \; \iff \; w <_{A_i} w'$$
that is, the ordering among $A$ valuations in $w$ and $w'$ is justified by the preference relation $<_A$; and vice-versa.
A weaker condition is faithfulness. A (multi-)preferential interpretation $\emme= \sx \WW, \{<_{A_i}\}, v \dx$ is {\em faithful} if, for all $w,w' \in \WW$, and preference relation $<_{A_i}$, 
$$ v(w,A_i) > v(w',A_i)  \; \Ri \; w <_{A_i} w'$$

Clearly, coherence is stronger than  faithfulness.
Furthermore, a preferential interpretation $\emme$ might be coherent with respect to a preference relation $<_{A_i}$, while being only faithful with respect to another $<_{A_j}$.

We can now define the satisfiability of a {\em graded implication} in  
a preferential interpretation $\emme= \sx \WW, \{<_{A_i}\}, v \dx$. 
\begin{definition}
A preferential interpretation $\emme= \sx \WW, \{<_{A_i}\}, v \dx$,
{\em  satisfies a graded implication $A  \rightarrow B \geq l$} (written $\emme \models A  \rightarrow B \geq l$) iff $(A  \rightarrow B)^\emme \geq l$. Similarly,
$I$ satisfies a graded implication $A  \rightarrow B \leq u$ (written $\emme \models A  \rightarrow B \leq u$) iff $(A  \rightarrow B)^\emme \leq u$.
\end{definition}

The satisfiability of a graded implication is evaluated globally to the preferential interpretation $\emme$.

Let a {\em knowledge base $K$} be a set of graded implications.
A {\em model of $K$} is an interpretation $\emme$ which satisfies all the graded implications in $K$.
Given a knowledge base $K$,  we say that $K$ {\em entails} a graded implication $A  \rightarrow B \geq l$ if $A  \rightarrow B \geq l$ is satisfied in all the models of $K$ (and similarly for a graded implication $A  \rightarrow B \leq l$). In the following, we will refer to the entailment of $A  \rightarrow B \geq 1$ as {\em $1$-entailment}.

Note that the two-valued case, with a single well-founded preference relation, can be regarded as a special case of this preferential logic, by letting ${\cal D}=\{0,1\}$, and assuming well-founded $<_A=<_B$, for all formulas $A$ and $B$.
In such a case, the faithful preferential semantics collapses to the usual KLM preferential semantics \cite{KrausLehmannMagidor:90}.

\subsection{KLM properties of conditionals}

The KLM properties of a {\em preferential consequence relation} can be reformulated in the many-valued setting, then proving that, for the choice of combination functions as in G\"odel logic, they
hold for 1-entailment.
Here, we assume ${\cal D}=[0,1]$ or  ${\cal D}={\cal C}_n$, for $n \geq 1$.

The KLM postulates of a preferential consequence relations \cite{KrausLehmannMagidor:90,whatdoes,Pearl:88} can be reformulated 
by replacing a conditional $A \ent B$ in the postulates with the conditional implication $\tip(A) \rightarrow B \geq 1$, as follows:

\medskip
$\mathit{\bf (Reflexivity)}$ \ $\tip(A) \rightarrow A \geq 1$ 

$\mathit{\bf (Left Logical Equivalence)}$ \ If 
$\models A \leftrightarrow  B$  and \\ $\tip(A) \rightarrow C \geq 1$, then $\tip(B) \rightarrow C  \geq 1$ 

$\mathit{\bf (Right Weakening)}$ \  If $\models B \rightarrow C$ and $\tip(A) \rightarrow B \geq 1$, then $\tip(A) \rightarrow C \geq 1$ 

$\mathit{\bf (And)}$ \ If $\tip(A) \rightarrow B \geq 1 $ and $\tip(A) \rightarrow C \geq 1$, then $\tip(A) \rightarrow B \wedge C \geq 1$

$\mathit{\bf (Or)}$ \ If $\tip(A) \rightarrow C \geq 1$ and $\tip(B) \rightarrow C \geq 1$, then $\tip(A \vee B) \rightarrow C \geq 1$

$\mathit{\bf (Cautious Monotonicity)}$ \  If $\tip(A) \rightarrow C \geq 1$ and $\tip(A) \rightarrow B \geq 1$, then $\tip(A \wedge B) \rightarrow C \geq 1$.
\medskip

Note that $A,B$ and $C$ above, do not contain the typicality operator.
Here, we also reinterpret $\models A \rightarrow  B$ as the requirement that $ A \rightarrow  B \geq 1$  
is satisfied in all many-valued interpretations, 
and that
$\models A\leftrightarrow B$ holds if  both $\models A \rightarrow  B$ and $\models B \rightarrow  A$ hold.

For the meaning of the postulates let us consider, for instance, the meaning of $\mathit{(Right Weakening)}$:
if  $\models B \rightarrow  C \geq 1$ holds,
and  $\tip(A) \rightarrow C \geq 1$ is entailed by a knowledge base $K$, then $\tip(A) \rightarrow B \geq 1$ is also entailed by the knowledge base $K$.

We can prove the following result.
\begin{proposition} \label{prop:KLM_properties}
Under the choice of combination functions as in G\"odel logic, 1-entailment
satisfies the KLM postulates of a preferential consequence relation given above.
\end{proposition}

This result for the many-valued propositional case, is the analogue of a similar result for many-valued, multi-preferential description logics $\alc$ with typicality \cite{IJAR23}.
Here, we are as well restricting to the truth valued set to ${\cal D}= [0,1]$ (or to finite subsets of interval $[0,1]$).

The KLM properties above do not exploit negation and 
they also hold for Zadeh's logic. Some of the properties above might not hold for other choices of combination functions (as in many-valued  DLs with typicality  \cite{IJAR23}). 
Note that whether the KLM properties are  intended or not, may depend on the kind of conditionals and on the kind of reasoning one aims at, which is still a matter of debate 
\cite{BonattiSauro17,KoutrasKR18,Rott2019,CasiniJelia2019}.


\section{A Temporal  Preferential Logic with Typicality} \label{sec:Conditional_reasoning}

In this section we extend the language of the logic ${\cal L}^\tip$ with the temporal operators $\bigcirc$ (next), $\Unt$ (until), $\Diamond$ (eventually) and $\Box$ (always) of Linear Time Temporal Logic (LTL) \cite{Clarke_book_1999}.

First we extend the language of graded implications, by allowing temporal and typicality operators to occur in a graded implication $A \ri B \geq l$ (or  $A \ri B \geq l$) in $A$ and in $B$, with the only restriction that $\tip$ should not be nested.
For instance, 

$\mathit{lives\_in\_town} \wedge \mathit{young} \ri \tip(\Diamond \mathit{granted\_loan}) \geq 0.8 $

\noindent
is a graded implication, as well as

$\Diamond \tip( \mathit{granted\_loan}) \ri \mathit{lives\_in\_town} \wedge \mathit{young}  \geq 0.8 $.

\noindent
We define the semantics of the 
logic in agreement with the fuzzy LTL semantics by Frigeri et al. \cite{Frigeri2014}.
\begin{definition}
A {\em temporal (multi-)preferential  interpretation}  is a triple $\cali= \sx \WW, \{<^n_{A_i}\}_{n \in \mathbb{N}} , v \dx$ where:
\begin{itemize} 
\item $\WW$ is a non-empty set of worlds;
\item  each $<^n_{A_i} \subseteq \WW \times \WW$ is an irreflexive, transitive and well-founded relation on $\WW$;
\item  $v:  \mathbb{N} \times  \WW \times \mathit{Prop} \longrightarrow {\cal D}$ is a valuation function assigning, at each time point, a truth value to any propositional variable in each world $w \in \WW$. 
\end{itemize}
\end{definition}
When there is no  $w' \in \WW$ s.t. $w' <_A^n w$, we say that $w$ is a normal situation for $A$ at timepoint $n$.

In a preferential interpretation $\cali= \sx \WW, \{<^n_{A_i}\}_{n \in \mathbb{N}} , v \dx$, the valuation $v(n,w,A)$ of a  formula $A$, in world $w$ at time point $n \in \mathbb{N}$, can be defined inductively as follows:

  $\mbox{\ \  }$ $v(n,w,\bot)=0_{\cal D}$ \ \ \ \ \ \ \ \ \  $v(n,w,\top)=1_{\cal D}$

  $\mbox{\ \  }$  $v(n,w, A \wedge B)= v(n,w,A) \otimes v(w,B)$
  
  $\mbox{\ \  }$ $v(n,w,A \vee B)= v(n,w,A) \oplus v(n,w,B)$ 
       
  $\mbox{\ \  }$ $v(n,w,\neg A)= \ominus v(n,w,A)$

  $\mbox{\ \  }$ $v(n, w, \tip(A))= v(n, w, A)$, if $\nexists w' \in \WW$ s.t. \\
  $\mbox{\ \ \ \ \ \ \ \ \ \ \ \ \ \ \ }$ \hspace{1.5cm} $w' <_A^n w$; \ \ $0_{\cal D}$ otherwise.

  $\mbox{\ \ }$   $v(n,w,\bigcirc A)= v(n+1,w,A)$
  
  $\mbox{\ \  }$    $v(n,w,\Diamond A)=  \bigoplus_{m \geq n} v(m,w,A) $
  
  $\mbox{\ \  }$    $v(n,w,\Box A)=  \bigotimes_{m \geq n} v(m,w,A) $

  $\mbox{\ \  }$   $v(n,w, A \Unt B) =  \bigoplus_{m \geq n} (v(m,w, B) \otimes $\\
  $\mbox{\ \  }$    \ \ \ \ \ \ \ \ \ \ \ \ \ \ \ \ \ \ \ \ \ \ \ \ \ \ \ \ \ \ \ \ \ \ \ \ \ \ \ \ \ \ \ \ \ \ \ \ $ \bigotimes_{k = n}^{m-1} v(k,w,A) )$

The semantics of $\Diamond$, $\Box$ and $\Unt$ requires a passage to the limit. 
Following  \cite{Frigeri2014}, we introduce a  bounded version for $\Diamond$, $\Box$ and $\Unt$, by adding new temporal operators  $\Diamond_t$ (eventually in the next $t$ time points), $\Box_t$ (always within $t$ time points) and $\Unt_t$, 
with the interpretation: 

$\mbox{\ \ \ }$  $v(n,w ,\Diamond_t A) =  \bigoplus_{m = n}^{n+t} v(m,w,A) $

$\mbox{\ \ \ }$  $v(n,w ,\Box_t A) =  \bigotimes_{m = n}^{n+t} v(m,w,A) $

$\mbox{\ \ \ }$  $v(n,w , A \Unt_t B) =  \bigoplus_{m = n}^{n+t}  (v(m,w,B) \otimes$ \\ 
$\mbox{\ \ \ }$  \ \ \ \ \ \ \ \ \ \ \ \ \ \ \ \ \ \ \ \ \ \ \ \ \ \  \ \ \ \ \ \ \ \ \ \ \   \ \ \ \ \ \ \ \ \ \ \ $\bigotimes_{k = n}^{m-1} v(k,w,A) )$

so that 

$\mbox{\ \ \ }$  $v(n,w,\Diamond A) = lim_{t \ri + \infty}  \; v(n,w ,\Diamond_t A) $ 

$\mbox{\ \ \ }$  $v(n,w,\Box A) = lim_{t \ri + \infty}  \; v(n,w ,\Box_t A) $ 
 
$\mbox{\ \ \ }$   $v(n,w, A \Unt B)  = lim_{t \ri + \infty} v(n,w , A \Unt_t B)$ 
  
\noindent
The existence of the limits is ensured by the fact that 
$( \Diamond C)^{\cal I}(n,x)$ and $(C \Unt D)^{\cal I}(n,x)$ are increasing in $n$, while $( \Box C)^{\cal I}(n,x) $ is decreasing in $n$  \cite{Frigeri2014}.

Note that, here, we have not considered the additional temporal operators (``soon'', ``almost always'', etc.)   
introduced by Frigeri et al. \cite{Frigeri2014}  for representing vagueness in the temporal dimension (which can be considered for future work).
As a consequence, for the case ${\cal D}=[0,1]$, without the typicality operator, the semantics corresponds to the semantics of FLTL (Fuzzy Linear-time Temporal Logic) by Lamine and Kabanza \cite{LamineKabanza2000}. 

\begin{proposition} For any formulas $A$ and $B$, and time point $n$, the following holds:

$v(n,w,\Diamond A) =  v(n,w,A)  \oplus v(n+1,w,\Diamond A) $

$v(n,w,\Box A) =  v(n,w,A)  \otimes v(n+1,w,\Box A) $

$v(n,w, A \Unt B) =  v(n,w,B)  \oplus$ \\
$\mbox{\ \ \ }$  \ \ \ \ \ \ \ \ \ \ \ \ \ \ \ \ \ \ \ \ \ \ \ \ \ \ \ \ $ ( v(n,w,A)  \otimes v(n+1,w, A \Unt B) )$

\end{proposition}

We can see that a temporal many-valued  interpretation $\cali= \sx \WW, \{<^n_{A_i}\}_{n \in \mathbb{N}} , v \dx$ can be regarded as a sequence 
of (non-temporal) preferential interpretations $\emme_0, \emme_1, \emme_2, \ldots$ where each $\emme_n$ is defined as follows: $\emme_n= \sx \WW, \{<^n_{A_i}\}, v^n \dx$, where
$w <^n_{A_i} w'$ holds in $\emme_n$ iff $w <^n_{A_i} w'$ holds in $\cali$, for all $w,w' \in \WW$; 
and
$v^n(w,A)=v(n,w,A)$, for all $w \in \WW$.

For the choice of ${\cal D}=[0,1]$, and of combination functions as in G\"odel logic, at each single time point the KLM properties of a preferential consequence relation are then expected to hold by Proposition \ref{prop:KLM_properties}.

\begin{definition}\label{def:degree_of_implication_n}
Given a temporal preferential interpretation  $\cali= \sx \WW, \{<^n_{A_i}\}_{n \in \mathbb{N}} , v \dx$ 
the {\em truth degree of an implication $A \rightarrow B$  in  $\cali$ at time point $n$} is defined as: \\ 
$\mbox{\ \ \ }$ $(A \rightarrow B)^{\cali,n}= inf_{w \in \WW} (v(n, w,A) \rhd v(n, w, B) )$.
\end{definition}

\medskip
Let us now define the {\em satisfiability of a  graded implication} in  
a preferential interpretation $\cali= \sx \WW, \{<^n_{A_i}\}_{n \in \mathbb{N}} , v \dx$. 

Rather than regarding graded implications as global constraints, that have to hold at all the time points,  we can allow for boolean combination of graded implications (as in \cite{NMR2023}) and also  for temporal operators to occur in front of the graded implications and of their boolean combinations. We call such formulas {temporal graded formulas}.


\subsection{Temporal graded Formulas} \label{sec:temp_graded_impl}

A {\em temporal graded formula} is defined as follows:

$\alpha::= A  \rightarrow B \geq l \mid A  \rightarrow B \geq l \mid \alpha \wedge \beta \mid \neg \alpha \mid$

$\mbox{\ \ \ \ }$ \ \ \ \ \ $ \bigcirc \alpha \mid \Diamond \alpha \mid \Box \alpha \mid \alpha \Unt \beta$,

\noindent
where $\alpha$ and $\beta$ stand for temporal graded formulas.
Note that temporal operators may occur both within graded implications ($A  \rightarrow B \geq l $) and in front of them, and of their boolean combinations.

An example of temporal graded formula is the following conjunction:

 $  \Box( \tip(\mathit{professor}) \ri  teaches \ \Unt \ \mathit{retired} \geq 0.7)$
    $\wedge$
 \\   
 ($\mathit{lives\_in\_town} \wedge \mathit{young} \ri \tip(\Diamond \mathit{granted\_loan}) \geq 0.8 $)
 
\noindent
where the graded implication in the first conjunct is prefixed by a $\Box$ operator, while the second one is not.

A {\em temporal conditional KB} is a set of temporal graded formulas.

We will evaluate the satisfiability of a temporal graded formula at the initial time point $0$ of a temporal preferential interpretation $\cali$.

Let us first define the interpretation of temporal graded formulas at a time point $n$ of a temporal interpretation $\cali$ as follows:

$\cali,n \models A  \rightarrow B \geq l$ iff $(A  \rightarrow B)^{\cali,n} \geq l$

$\cali,n \models A  \rightarrow B \leq l$ \ iff \ $(A  \rightarrow B)^{\cali,n} \leq l$

$\cali,n \models \alpha  \wedge \beta$\  iff \ $\cali,n \models \alpha  $ and $\cali,n \models \beta$

$\cali,n \models \neg \alpha $\  iff \  $\cali,n \not \models \alpha  $ 

$\cali,n \models \bigcirc \alpha $\  iff \ $\cali,n+1 \models \alpha  $ 

$\cali,n \models \Diamond \alpha $\  iff \ exists $m \geq n$ such that  $\cali,m \models \alpha  $

$\cali,n \models \Box \alpha $\  iff \ for all $m \geq n$,  $\cali,m \models \alpha  $

$\cali,n \models \alpha \Unt \beta $\  iff \ exists $ m \geq n$ such that  $\cali,m \models \beta  $ 

$\mbox{\ \ \ \ }$ \ \ \ \ \ \ \ \ \ \ \ \ \ \ \ \ \ \ \ \ \ \ \  and, for all $n \leq k < m  $, \  $\cali,k \models \alpha  $ 

\noindent
Let us define the notions of satisfiability and entailment.
\begin{definition}[Satisfiability and entailment]\label{satisfiability_temp}
A graded formula $\alpha$ is {\em  satisfied} in a temporal preferential interpretation $\cali= \sx \WW, \{<^n_{A_i}\}_{n \in \mathbb{N}} , v \dx$ \  if \ 
$\cali,0 \models \alpha$.

A  preferential interpretation $\cali= \sx \WW, \{<^n_{A_i}\}_{n \in \mathbb{N}} , v \dx$ is a {\em model} of a temporal conditional knowledge base $K$, if $\cali$ satisfies all the temporal graded formulas in $K$.

A temporal conditional knowledge base $K$ {\em entails} a temporal graded formula $\alpha$ if $\alpha$ is satisfied in all the models $\cali$ of $K$.
\end{definition}
Observe that any graded implication $A  \rightarrow B \geq l$ is either satisfied or not  at a time point  $n$ of a temporal interpretation $\cali$, i.e., either $\cali, n \models A  \rightarrow B \geq l$ or $\cali, n \not \models A  \rightarrow B \geq l$ (and similarly for the graded implications with $\leq$). Hence, the interpretation above of temporalized formulas in $\cali$ at a time point $n$  is two-valued (although it builds over the degree of an implication $A \ri B$ in $\cali$ at time point $n$, which has a truth value  $(A  \rightarrow B)^{\cali,n}$ in ${\cal D}$, see Definition \ref{def:degree_of_implication_n}).

Note that, in the temporal graded formula given above, the graded implication in the first conjunct ($\tip(\mathit{professor}) \ri  teaches \ \Unt \ \mathit{retired} \geq 0.7$) is required to hold at all the time points of the interpretation $\cali$ (as it is prefixed by $\Box$), while the second conjunct ($\mathit{lives\_in\_town} \wedge \mathit{young} \ri \tip(\Diamond \mathit{granted\_loan}) \geq 0.8 $) has to hold only at time point $0$.

Decidability and complexity of the different decision problems (the satisfiability, the model checking and entailment problems) have to be studied for this temporal many-valued conditional logic, for different choices of ${\cal D}$ and of the combination functions.
Satisfiability is decidable in the two-valued case, when we restrict to preference relations $<_{A_i}$ with respect to a finite number of formulas (for instance,  by restricting to the formulas occurring in a finite KB, and to the respective preferences). Under such conditions,  the  propositional temporal logic with typicality introduced above can be regarded as a special case of $LTL_\alc$ with typicality, which has been shown to be decidable in \cite{CILC2023} for a finite number of preference relations.

\section{Weighted temporal knowledge bases} \label{sec:weighted_KBs}

As in the two-valued non-temporal case, the notion of preferential entailment considered in the previous section is rather weak.
For the KLM logics, some different  closure constructions have been proposed to strengthen entailment by restricting to a subset of the preferential models of a conditional knowledge base $K$. Let us just mention, the rational closure \cite{whatdoes} (or system Z \cite{Pearl:88}) and the lexicographic closure \cite{Lehmann95}, 
but also other constructions, such as the MP-closure \cite{AIJ21}, which exploit a similar idea, but using a different kind of lexicographic ordering to define the preference relation.

In the following we consider a construction that has been proposed for weighted knowledge bases in defeasible description logics, where defeasible implications have a weight.
 We reformulate the semantics in a propositional context, for the temporal case, by assuming that ${\cal D}$ is the unit interval $[0,1]$ or a subset of it (e.g., the finite set  ${\cal D}={\cal C}_n$, for some $n \geq 1$). The two-valued case ${\cal D}=\{0,1\}$ is also a special case.

A  {\em weighted KB}  is a set of 
\emph{weighted typicality implication} of the form $\left(\tip(A_i) \ri B_j, w_{ij}\right)$, 
where $A_i$ and $B_j$ are propositions, and the weight $w_{ij}$ is a real number, representing the  plausibility or implausibility of  the conditional implication.
For instance, for a proposition $\mathit{student}$, 
we may have a set of weighted defeasible implications:

\ \ ($\mathit{\tip(student) \ri has\_Classes}$, +50), 

\ \ ($\mathit{\tip(student) \ri  \Diamond holds\_Degree}$,+30) ,

\ \ ($\mathit{\tip(student) \ri  has\_Boss}$, -40),

\noindent
that represent {\em prototypical properties} of students, i.e., that a student normally has classes and will eventually reach the degree, but she usually does not have a boss (negative weight). 
Accordingly, a student having classes, but not a boss, is more typical than a student having classes and a boss.
Similarly, one may introduce a set of weighted conditionals for other propositions, e.g., for $\mathit{employee}$.

Based on the set of conditionals, one can establish the preferences between the different worlds with respect to different propositions $A_i$.
For instance, consider an interpretation $\cali= \sx \WW, \{<^n_{A_i}\}_{n \in \mathbb{N}} , v \dx$ in which a world $w$ describes a student ($v(0,w,\mathit{student})=1$) that in the initial state has classes ($v(0,w,\mathit{has\_Classes})=1$) but not a boss ($v(0,w,\mathit{has\_Boss})=0$), and that at time point 8 will reach the degree ($v(8,w,\mathit{hold\_Degree})=1$) ; while world $w'$ describes a student $v(0,w',\mathit{student})=1$ that in the initial state has classes ($v(0,w',\mathit{has\_Classes})=1$)  and has a boss ($v(0,w',\mathit{has\_Boss})=1$), and will reach the degree at time point $7$ ($v(7,w',\mathit{hold\_Degree})=1$.

The idea is that the preference relation $<_\mathit{student}$ in $\cali$ should consider the situation described at $w$ at time point $0$, more normal than the situation described by $w'$, i.e.,  $w <_\mathit{student} w'$, as the sum of the weights of the defeasible implications satisfied by world $w$ ($50+30=80$) is greater than the sum of the  weights of the defeasible implications satisfied by world $w'$  ($50+30-40=40$). 

We have to further consider that the propositions may be non-crisp, e.g., 
$v(0,w,\mathit{has\_Classes})=0.7$, and this has  some impact on the degree to which a conditional implication (e.g., $\mathit{\tip(student) \ri has\_Classes}$),  is satisfied.

Given a weighted knowledge base $K$, we call {\em distinguished propositions} those propositions $A_i$ such that at least a weighted defeasible implications of the form $\left(\tip(A_i) \ri B_j, w_{ij}\right)$ occurs in $K$.

Let $K$ be a temporal weighted KB.
Given a many-valued temporal interpretation $\cali= \sx \WW, \{<^n_{A_i}\}_{n \in \mathbb{N}} , v \dx$, the \emph{weight of a world $x \in \WW$ with respect to a distinguished proposition $A_i$ at time point $n$} is given by
\begin{center}
$W^{\cal I}_{{A_i},n}(x) = \sum_{\left(\tip(A_i) \ri B_j, w_{ij}\right) \in {K}}\;  { \;  w_{ij} \cdot v(n,x,B_j)}. $

\end{center}
Intuitively, the higher the value of $W^{\cal I}_{i,n}(x)$, the more normal is  the state of affairs $x$, at time point $n$, concerning the properties of $A$ in $K$. 

\begin{definition}\label{defi:preferences}
A many-valued temporal preferential interpretation $\cali= \sx \WW, \{<^n_{A_i}\}_{n \in \mathbb{N}} , v \dx$ satisfies a weighted KB $K$ if, for all the distinguished formulas $A_i$, it holds that:
\begin{center}
$x <^n_{A_i }y $ \ \ \ \ \ $\iff$ \ \ \ \ \  $W^{\cal I}_{i,n}(x) > W^{\cal I}_{i,n}(y)$
\end{center}
\end{definition}
The condition in Definition \ref{defi:preferences}, together with the coherence (faithfulness) condition introduced in Section \ref{sec:many_valued_LTL}, guarantees that the many-valued interpretation $\cali$ agrees with the weighted inclusions in $K$, at each time point $n$.

A weighted (defeasible) knowledge base $K_D$ can coexist with a strict knowledge base $K_S$ (i.e., a set of graded implications).
This is the usual approach in defeasible DLs.


\section{Towards a temporal conditional logic for gradual argumentation}\label{sec:gradual_argumentation}

In previous sections, we have developed a many-valued, temporal logic with typicality, extending with $LTL$ operators the many-valued conditional logic with typicality proposed in \cite{NMR2023}. 
In this section we aim at instantiating the proposed temporal logic to the gradual argumentation setting, to make it suitable for capturing the dynamics of an argumentation graph (e.g., the changes of weights of edges in time). 

The idea in \cite{NMR2023} was to provide a general approach for developing  a preferential interpretation  from an argumentation graph $G$ under a gradual semantics $S$, provided some weak conditions on the domain of argument interpretation are satisfied and, specifically, that the {\em  domain of argument interpretation} ${\cal D}$ is equipped with a {\em preorder relation $\leq$} (which is a widely agreed requirement  \cite{BaroniRagoToni2018,BaroniRT19}).
As it may be expected, the domain of argument interpretation ${\cal D}$ plays the role of the truth degree set of our many-valued semantics introduced above.

For the definition of an argumentation graph, let us adapt the notion of {\em edge-weighted QBAF} by  Potyka \cite{PotykaAAAI21} to a generic domain ${\cal D}$.
A {\em (weighted) argumentation graph} is a quadruple
$G=\la \mathcal{ A}, \mathcal{ R}, \sigma_0, \pi \ra$, where $\mathcal{ A}$ is a set of arguments, $\mathcal{ R} \subseteq \mathcal{ A} \times \mathcal{ A}$ a set of edges, $\sigma_0: \mathcal{ A} \rightarrow {\cal D}$ assigns a  {\em base score} of arguments,
and $\pi: \mathcal{ R} \rightarrow  \mathbb{R}$ is a weight function assigning a positive or negative weight to edges.

A {\em  labelling $\sigma$ of $G$ over ${\cal D}$} is a function $\sigma: {\cal  A} \rightarrow {\cal D}$, 
which assigns to each argument an {\em acceptability degree} (or a {\em strength}) in the domain of argument valuation $ {\cal D} $.
Whatever semantics $S$ is considered for an argumentation graph $G$, we assume that $S$ identifies {\em a set $\Sigma^S$ of labellings} of the graph $G$ over a domain of argument valuation ${\cal D}$. 

A {\em semantics $S$ of $G$} can then be regarded, abstractly, as a pair $({\cal D},\Sigma^S)$: a domain of argument valuation $ {\cal D} $ and a set of labellings $\Sigma^S$ over the domain. 

If we consider all arguments $A_i \in {\cal A}$ as  propositional variables, 
each labelling $\sigma$ can be regarded as a world $w_\sigma \in \WW$ in a many-valued preferential interpretation $\emme^G= \sx \WW, \{<_{A_i}\}, v \dx$, such that $v(w_\sigma, A_i)= \sigma(A_i)$.

More precisely, a gradual  semantics $({\cal D},\Sigma^S)$ of the argumentation graph $G$ can be mapped into a preferential interpretation
$\emme^G= \sx \WW, \{<_{A_i}\}, v \dx$, defined as  in Section \ref{sec:many_valued_LTL}, by letting:

- $\WW=\{ w_\sigma \mid \sigma \in \Sigma^S\}$

- $v(w_\sigma, A_i)= \sigma(A_i)$, for all the arguments $A_i \in Prop$ 

- $w_\sigma <_{A_i} w_{\sigma'}$ \ iff \ $\sigma(A_i) >\sigma'(A_i)$

\noindent
Such a preferential interpretation can then be used in the verification of strict and conditional graded implications. 
For a specific gradual argumentation semantics, in the finitely-valued case, an ASP approach for conditional reasoning over an argumentation graph, has been presented in \cite{ASPOCP23}.

The approach can be extended to the temporal case, based on the temporal many-valued logic with typicality developed in Section  \ref{sec:Conditional_reasoning}. 

It allows to reason about the dynamics of an argumentation graph, when the weights of edges might change in time, e.g. when learning the weights.
Indeed, a multilayer neural network can be regarded as an argumentation graph   \cite{PotykaAAAI21,GarcezBG01}, or as a weighted knowledge base \cite{WorkshopAI3,IJAR23}, based on the strong relationships of the two formalisms \cite{Intell_Artif_2024}.
As another example, the structure of the argumentation graph can be updated through the interaction of different agents in time, such as in  \cite{RagoLiToniKR23} via Argumentative Exchanges. 

The labellings of the graph at different time points can be used for constructing a temporal interpretation $\cali$ as a sequence of non-temporal interpretations $\emme_0, \emme_1, \ldots$ (as $\emme^G$ above), and temporal graded formulas over arguments, e.g.,
$  \Box( \tip(\mathit{A_1}) \ri  A_2 \Unt \mathit{A_3} \vee A_3 ) \geq 0.7$,
can be verified over $\cali$.

As mentioned above,
this verification approach has been studied, for the non-temporal case, in the verification of properties of  argumentation graphs under the $\varphi$-coherent gradual semantics  \cite{ASPOCP23}, and an ASP approach has been developed for the verification of graded conditional implications over arguments and over boolean combination of arguments.
Extending the ASP approach to deal with the temporal case, for specific fragments of the language, is a direction for future work.

\section{Conclusions}
\label{sec:conclusions}

The paper proposes a framework in which different (many-valued) preferential logics with typicality can be captured, together with their temporal extensions, with the operators from LTL. 
The interpretation of the typicality operator is based on a multi-preferential semantics, and an extension of weighted conditional  knowledge bases  to the temporal (many-valued) case is proposed.

The approach is parametric with respect to the choice of a specific many-valued logic (with their combination functions), but also with respect to the definition of the preference relations $<_{A_i}$, which may exploit different closure constructions, among the many studied in the literature, in the spirit of Lehmann's lexicographic closure \cite{Lehmann95}.
The two-valued case, with a single preference relation 
can as well be regarded as a special cases of this preferential temporal formalism.

On a different route, a preferential logics with defeasible LTL operators has been studied in \cite{ChafikTIME2020,ChafikJANCL23}. The decidability of different fragments of the logic has been proven, and tableaux based proof methods for such fragments have been developed \cite{ChafikACV21,ChafikJANCL23}. Our approach does not consider defeasible temporal operators nor preferences over time points, but combines standard LTL operators with the typicality operator in a many-valued temporal logic. 
In our approach, preferences between worlds change over time.

Much work has been recently devoted to the combination 
of neural networks and symbolic reasoning \cite{SerafiniG16,GarcezGori2020,Guidotti21}.
While conditional weighted KBs have been shown to capture (in the many-valued case) the stationary states of a  
neural network (or its finite approximation) \cite{JELIA2021,IJAR23,Intell_Artif_2024},
and allow for combining empirical knowledge with elicited knowledge for reasoning and for post-hoc verification,
adding a temporal dimension opens to the possibility of verifying properties concerning the dynamic behaviour of the network, based on a model checking approach 
or an entailment based approach.

Extending the above mentioned ASP encodings to deal with temporal preferential interpretations is a direction of future work.
Future work also includes studying the decidability for fragments of the logic 
and  exploiting the formalism for explainability, 
and for reasoning about the dynamics of gradual argumentation graphs in gradual semantics.

\end{document}